# Instantaneous Relaying for the 3-Way Relay Channel with Circular Message Exchanges

Bho Matthiesen and Eduard A. Jorswieck
Communications Theory, Communications Laboratory
Department of Electrical Engineering and Information Technology
Technische Universität Dresden, Germany
Email: {bho.matthiesen, eduard.jorswieck}@tu-dresden.de

*Abstract*—The 3-user discrete memoryless multi-way relay channel with circular message exchange and instantaneous relaying is investigated. We first show that this channel is effectively a 3-user interference channel with receiver message side information for every fixed (and instantaneous) relay mapping. Then, we extend the Han-Kobayashi coding scheme to this channel. Finally, we apply these results to Gaussian channels with amplify-and-forward relaying and present numerical results showing the gain of the proposed scheme compared to the state of the art.

## I. INTRODUCTION

We study the 3-user discrete memoryless multi-way relay channel (DM-MWRC) with circular message exchange and instantaneous relaying. The MWRC models relay-aided communication across several nodes with no direct links between the users. Applications of this model are, for example, communication of several ground stations over a satellite, wireless board-to-board communication in highly adaptive computing [1] where multiple chips exchange data with the help of another chip acting as relay, heterogeneous dense small cell networks in modern and future 5G wireless networks, Industry 4.0, or wireless sensor networks.

Instantaneous relaying [2] restricts the relay to operate deterministically, memoryless and symbol-wise. That is, the current output signal depends only on the currently observed input symbol. A famous example for instantaneous relaying is the linear amplify-and-forward (AF) relaying scheme [3], [4]. However, it is shown in [2] that AF is not the optimal instantaneous relaying scheme for the Gaussian relay channel with orthogonal receive components. Restricting the relay to operate instantaneously has two main reasons: First, the processing delay at the relay is significantly less than in other schemes like decode-and-forward (DF). Second, the relay does not need power hungry analog-to-digital conversation and digital signal processing which results in reduced hardware complexity and better energy efficiency (EE). The downside is that noise is not cancelled at the relay. However, it is shown in [5] that in terms of EE AF can outperform DF and similar schemes due to reduced circuit power. These advantages make it a favorable choice for highly adaptive energy efficient computing or sensor networks.

The Gaussian MWRC is studied in [6] as a model for clustered communication over a relay, where the terminals in each cluster exchange information among each other with the help of a relay. They require each user to decode all other messages sent by users in the same cluster and present achievable rate regions for AF, DF, compress-and-forward (CF), and lattice codes. Most other works on MWRCs consider the case with only one cluster. An extensive overview of results for this channel is given in [7].

We focus on the general class of DM-MWRCs with instantaneous relaying. Due to the circular message exchange, each node only wants to recover one message and, hence, has to deal with interference. Some of this interference is self-induced and can be removed, but the remaining interference needs to be addressed. Common approaches are treating interference as noise (IAN) or performing simultaneous non-unique decoding (SND). However, with rate splitting and superposition coding a combination of these two techniques is possible which, in general, results in higher transmission rates. This coding scheme was first introduced by Han and Kobayashi [8] for the 2-user interference channel (IC). We first show that the DM-MWRC with instantaneous relaying is, for every fixed relaying function, equivalent to a 3-user IC with receiver message side information and feedback. Then, we extend the Han-Kobayashi (HK) coding scheme to this channel and derive an achievable rate region, which, to the best of our knowledge, is the largest achievable rate region for this channel known to date. From this result, we derive achievable rate regions using only SND or IAN as corollaries. Finally, we extend the results to Gaussian channels with AF and provide numeric evidence that HK coding achieves higher sum rates than SND and IAN.

Throughout this paper, we use the same notation as in [9].

## II. SYSTEM MODEL

We consider the 3-user DM-MWRC with partial message exchange illustrated in Fig. 1. The considered message exchange has two defining properties:

1) Each user has a message to transmit which is intended for at least one other user.

This work is supported by the German Research Foundation (DFG) in the Collaborative Research Center 912 "Highly Adaptive Energy-Efficient Computing."

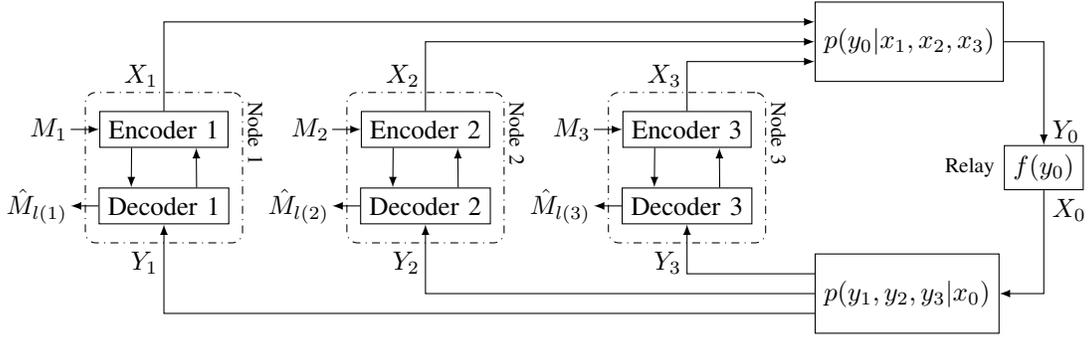

Fig. 1. Block diagram of the 3-user DM-MWRC with circular message exchange and instantaneous relaying.

2) Each user desires at most one message.

From these two properties it follows immediately that each message is only required at one other user. We denote the message of user $k$, $k \in \mathcal{K} = \{1, 2, 3\}$, as $M_k$ and the node receiving it as $q(k)$. Furthermore, the user not interested in $M_k$ is denoted by $l(k)$. Also, it follows from the properties listed above, that user $k$, $k \in \mathcal{K}$, desires the message sent by user $l(k)$. Further, since $M_k$ and $M_{l(k)}$ are to be decoded by users $q(k)$ and $k$, respectively, $M_{q(k)}$ is the desired message at node $l(k)$. This leaves us with two possible message exchanges, either clockwise or counter-clockwise, depending on whether $q(1)$ is 2 or 3, respectively. Furthermore, it follows immediately that $q(q(k)) = l(k)$, $l(q(k)) = k$, $q(l(k)) = k$, $l(l(k)) = q(k)$, $q^{-1}(k) = l(k)$, and $l^{-1}(k) = q(k)$.

Finally, we assume the messages to be independent. They are communicated in $n$ channel uses with the help of a relay. Each message $M_k$, $k \in \mathcal{K}$, is encoded into a codeword $X_k^n$ of length $n$ and transmitted over the channel. Upon receiving $Y_{q(k)}^n$, receiver $q(k)$, $k \in \mathcal{K}$, finds an estimate $\hat{M}_k$ of message $M_k$ using its own message $M_{q(k)}$ as side information. We assume full-duplex operation at all nodes.

The 3-user DM-MWRC ($\mathcal{X}_0 \times \mathcal{X}_1 \times \mathcal{X}_2 \times \mathcal{X}_3, p(y_0, y_1, y_2, y_3 | x_0, x_1, x_3, x_4), \mathcal{Y}_0 \times \mathcal{Y}_1 \times \mathcal{Y}_2 \times \mathcal{Y}_3$) consists of four finite input sets $\mathcal{X}_0, \mathcal{X}_1, \mathcal{X}_2, \mathcal{X}_3$, four finite output sets $\mathcal{Y}_0, \mathcal{Y}_1, \mathcal{Y}_2, \mathcal{Y}_3$, and a collection of conditional probability mass functions (pmfs) $p(y_0, y_1, y_2, y_3 | x_0, x_1, x_3, x_4)$ on $\mathcal{Y}_0 \times \mathcal{Y}_1 \times \mathcal{Y}_2 \times \mathcal{Y}_3$.

A $(2^{nR_1}, 2^{nR_2}, 2^{nR_3}, n)$ code for the 3-user DM-MWRC consists of

- three message sets $\mathcal{M}_k = [1 : 2^{nR_k}]$, one for each user $k \in \mathcal{K}$,
- three encoders, where encoder $k \in \mathcal{K}$ assigns a symbol $x_{ki}(m_k, y_k^{i-1})$ to each message $m_k \in \mathcal{M}_k$ and received sequence $y_k^{i-1}$ for $i \in [1 : n]$,
- a relay encoder that assigns a symbol $x_{0i}(y_0^{i-1})$ to every past received sequence $y_0^{i-1}$ for $i \in [1 : n]$, and
- three decoders, where decoder $q(k) \in \mathcal{K}$ assigns an estimate $\hat{m}_k \in \mathcal{M}_k$ or an error message $e$ to each pair $(m_{q(k)}, y_{q(k)}^n)$.

We assume that the message triple $(M_1, M_2, M_3)$ is uniformly distributed over $\mathcal{M}_1 \times \mathcal{M}_2 \times \mathcal{M}_3$. The average probability of error is defined as $P_e^{(n)} = \Pr\{\hat{M}_k \neq M_k \text{ for some } k \in \mathcal{K}\}$. A rate triple $(R_1, R_2, R_3)$ is said to be achievable if there exists a sequence of $(2^{nR_1}, 2^{nR_2}, 2^{nR_3}, n)$ codes such that $\lim_{n \to \infty} P_e^{(n)} = 0$. The capacity region of the 3-user DM-MWRC is the closure of the set of achievable rates.

In our model we assume no direct user-to-user links which implies that the channel decomposes into an upstream channel $p(y_0 | x_1, x_2, x_3)$ from the users to the relay, and a downstream channel $p(y_1, y_2, y_3 | x_0)$ from the relay back to the users. Further, we constrain the relay to operate instantaneously on the received symbol $y_{0i}$ using a deterministic mapping

$$f : \mathcal{Y}_0 \mapsto \mathcal{X}_0$$
$$x_{0i} = f(y_{0i}), \quad i = 1, \ldots, n.$$

Then, for every fixed $f$, the channel effectively is a 3-user discrete memoryless interference channel (DM-IC) with receiver message side information and causal feedback with inputs $(x_1, x_2, x_3)$, outputs $(y_1, y_2, y_3)$ and a collection of conditional pmfs $p(y_1, y_2, y_3 | x_1, x_2, x_3)$ given as

$$p(y_1, y_2, y_3 | x_1, x_2, x_3) = \sum_{y_0 \in \mathcal{Y}_0} p(y_0 | x_1, x_2, x_3) \, p_{Y_1 Y_2 Y_3 | X_0}(y_1, y_2, y_3 | f(y_0)). \quad (1)$$

Now, we can carefully adapt results for the classical DM-IC to our channel with receiver message side information.

## III. MAIN RESULTS & PROOFS

In Theorem 1 we present an achievable rate region for the DM-MWRC with instantaneous relaying using an HK [8] inspired coding scheme. It uses rate splitting to represent each message $M_k$, $k \in \mathcal{K}$, by an independent common message $M_k^c$ at rate $R_k^c$ and a private message $M_k^p$ at rate $R_k^p$. These messages are sent via superposition coding with cloud center $U_k(M_k^c)$ and satellite codeword $X_k(M_k^c, M_k^p)$. Receiver $q(k)$, $k \in \mathcal{K}$, recovers all common messages and its desired private message $M_k^p$ using its own message as side information. This approach not only incorporates IAN, i.e., $R_k^c = 0$ or $U_k = \emptyset$, and SND, i.e., $R_k^c = 0$ or $U_k = X_k$, but also allows for arbitrary combinations of those strategies. Furthermore, we present corollaries with achievable rate regions using IAN and

SND. For notational convenience, we define $R_\Sigma = R_1 + R_2 + R_3$.

*Theorem 1:* A rate triple $(R_1, R_2, R_3)$ is achievable for the DM-MWRC $p(y_0, y_1, y_2, y_3 | x_0, x_1, x_3, x_4)$ if, for all $k \in \mathcal{K}$,

$$R_k < B_k,$$
$$R_k + R_{q(k)} < A_k + D_{q(k)},$$
$$R_\Sigma < A_k + C_{q(k)} + D_{l(k)},$$
$$R_k + R_\Sigma < A_k + C_{q(k)} + C_{l(k)} + D_k,$$

and,
$$R_\Sigma < C_1 + C_2 + C_3,$$

where
$$A_k = I(X_k; Y_{q(k)} | U_k, U_{l(k)}, X_{q(k)}, Q) \quad (2)$$
$$B_k = I(X_k; Y_{q(k)} | U_{l(k)}, X_{q(k)}, Q) \quad (3)$$
$$C_k = I(X_k, U_{l(k)}; Y_{q(k)} | U_k, X_{q(k)}, Q) \quad (4)$$
$$D_k = I(X_k, U_{l(k)}; Y_{q(k)} | X_{q(k)}, Q) \quad (5)$$

for some pmf $p(q)p(u_1, x_1|q)p(u_2, x_2|q)p(u_3, x_3|q)$, and some deterministic mapping $f(y_0)$ at the relay, where $|\mathcal{U}_k| \leq |\mathcal{X}_k| + 3$, $k \in \mathcal{K}$, and $|\mathcal{Q}| \leq 9$.

*Corollary 1 (Treating interference as noise):* A rate triple $(R_1, R_2, R_3)$ is achievable for the DM-MWRC $p(y_0, y_1, y_2, y_3 | x_0, x_1, x_3, x_4)$ using IAN if, for all $k \in \mathcal{K}$,

$$R_k < I(X_k; Y_{q(k)} | X_{q(k)}, Q)$$

for some pmf $p(q)p(x_1|q)p(x_2|q)p(x_3|q)$, and some deterministic mapping $f(y_0)$ at the relay, where $|\mathcal{Q}| \leq 3$.

*Proof:* For all $k \in \mathcal{K}$, set $U_k = \emptyset$ in Theorem 1. The cardinality bound on $\mathcal{Q}$ follows from a standard Carathéodory type argument [9, Appendix C]. ∎

*Corollary 2 (Simultaneous non-unique decoding):* A rate triple $(R_1, R_2, R_3)$ is achievable for the DM-MWRC $p(y_0, y_1, y_2, y_3 | x_0, x_1, x_3, x_4)$ using SND if, for all $k \in \mathcal{K}$,

$$R_k < I(X_k; Y_{q(k)} | X_{l(k)}, X_{q(k)}, Q),$$
$$R_k + R_{l(k)} < I(X_k, X_{l(k)}; Y_{q(k)} | X_{q(k)}, Q),$$

for some pmf $p(q)p(x_1|q)p(x_2|q)p(x_3|q)$, and some deterministic mapping $f(y_0)$ at the relay, where $|\mathcal{Q}| \leq 6$.

The proof of this corollary is given in Section III-B.

Before we proof Theorem 1, we need the following lemma.

*Lemma 1:* A rate triple $(R_1, R_2, R_3)$ is achievable for the DM-MWRC $p(y_0, y_1, y_2, y_3 | x_0, x_1, x_3, x_4)$ if, for all $k \in \mathcal{K}$,

$$R_k < \min\{B_k, A_k + C_{q(k)}\}$$
$$R_k + R_{q(k)} < A_k + \min\{D_{q(k)}, C_{q(k)} + C_{l(k)}\},$$
$$R_\Sigma < A_k + C_{q(k)} + D_{l(k)},$$
$$R_k + R_\Sigma < A_k + C_{q(k)} + C_{l(k)} + D_k,$$

and,
$$R_\Sigma < C_1 + C_2 + C_3,$$

for some pmf $p(q)p(u_1, x_1|q)p(u_2, x_2|q)p(u_3, x_3|q)$, and some deterministic mapping $f(y_0)$ at the relay, where $A_k, \ldots, D_k$ are as defined in Theorem 1.

*Proof: Codebook generation:* Fix $p(q) \prod_{k \in \mathcal{K}} p(u_k, x_k | q)$ and $f(y_0)$. Generate a sequence $q^n$ according to $\prod_{i=1}^n p_Q(q_i)$. For each $k \in \mathcal{K}$, randomly and conditionally independently generate $2^{nR_k^c}$ sequences $u_k^n(m_k^c)$, $m_k^c \in [1 : 2^{nR_k^c}]$, each according to $\prod_{i=1}^n p_{U_k|Q}(u_{ki}|q_i)$. For each $m_k^c$, randomly and conditionally independently generate $2^{nR_k^p}$ sequences $x_k^n(m_k^c, m_k^p)$, $m_k^p \in [1 : 2^{nR_k^p}]$, each according to $\prod_{i=1}^n p_{X_k|U_k,Q}(x_{ki}|u_{ki}(m_k^c), q_i)$.

*Encoding:* To send $m_k = (m_k^c, m_k^p)$, encoder $k \in \mathcal{K}$ transmits $x_k^n(m_k^c, m_k^p)$.

*Decoding:* We use SND. Suppose that receiver $q(k)$ observes $y_{q(k)}^n$. Then it finds the unique $(\hat{m}_k^c, \hat{m}_k^p)$ such that

$$\bigl(q^n, u_k^n(\hat{m}_k^c), u_{l(k)}^n(m_{l(k)}^c), x_k^n(\hat{m}_k^c, \hat{m}_k^p),$$
$$u_{q(k)}^n(m_{q(k)}^c), x_{q(k)}^n(m_{q(k)}^c, m_{q(k)}^p), y_{q(k)}^n\bigr) \in \mathcal{T}_\epsilon^{(n)}$$

for some $m_{l(k)}^c \in [1 : 2^{nR_{l(k)}^c}]$ using its own messages $(m_{q(k)}^c, m_{q(k)}^p)$ as side information. Otherwise it declares an error.

*Analysis of the probability of error:* To bound the average probability of error for decoder $q(k)$, $k \in \mathcal{K}$, we assume without loss of generality that $(M_k^c, M_k^p) = (1, 1)$ is sent for all $k \in \mathcal{K}$. Then, the decoder makes an error if and only if one or more of the following events occur:

$$\mathcal{E}_{q(k),0} = \{(Q^n, U_k^n(1), U_{l(k)}^n(1), X_k^n(1,1), U_{q(k)}^n(1),$$
$$X_{q(k)}^n(1,1), Y_{q(k)}^n) \notin \mathcal{T}_\epsilon^{(n)}\},$$

$$\mathcal{E}_{q(k),1} = \{(Q^n, U_k^n(1), U_{l(k)}^n(1), X_k^n(1, m_k^p), U_{q(k)}^n(1),$$
$$X_{q(k)}^n(1,1), Y_{q(k)}^n) \in \mathcal{T}_\epsilon^{(n)} \text{ for some } m_k^p \neq 1\},$$

$$\mathcal{E}_{q(k),2} = \{(Q^n, U_k^n(m_k^c), U_{l(k)}^n(1), X_k^n(m_k^c, m_k^p),$$
$$U_{q(k)}^n(1), X_{q(k)}^n(1,1), Y_{q(k)}^n) \in \mathcal{T}_\epsilon^{(n)}$$
$$\text{ for some } m_k^c \neq 1, m_k^p\},$$

$$\mathcal{E}_{q(k),3} = \{(Q^n, U_k^n(1), U_{l(k)}^n(m_{l(k)}^c), X_k^n(1, m_k^p),$$
$$U_{q(k)}^n(1), X_{q(k)}^n(1,1), Y_{q(k)}^n) \in \mathcal{T}_\epsilon^{(n)}$$
$$\text{ for some } m_{l(k)}^c \neq 1, m_k^p \neq 1\},$$

$$\mathcal{E}_{q(k),4} = \{(Q^n, U_k^n(m_k^c), U_{l(k)}^n(m_{l(k)}^c), X_k^n(m_k^c, m_k^p),$$
$$U_{q(k)}^n(1), X_{q(k)}^n(1,1), Y_{q(k)}^n) \in \mathcal{T}_\epsilon^{(n)}$$
$$\text{ for some } m_k^c \neq 1, m_{l(k)}^c \neq 1, m_k^p\}.$$

Due to the union bound, the average probability of error for decoder $q(k)$ is bounded above as $\Pr(\mathcal{E}_{q(k)}) \leq \sum_{i=0}^4 \Pr(\mathcal{E}_{q(k),i})$.

By the weak law of large numbers, $\Pr(\mathcal{E}_{q(k),0})$ tends to zero as $n \to 0$. By the packing lemma [9] and since

$$U_k - (Q, X_k) - (Y_1, Y_2, Y_3, U_{q(k)}, X_{q(k)}, U_{l(k)}, X_{l(k)}) \quad (6)$$

form a Markov chain for every $k \in \mathcal{K}$, $\Pr(\mathcal{E}_{q(k),1})$, $\Pr(\mathcal{E}_{q(k),2})$, $\Pr(\mathcal{E}_{q(k),3})$, and $\Pr(\mathcal{E}_{q(k),4})$ tend to zero as $n \to 0$ if the conditions $R_k^p < I(X_k; Y_{q(k)} | U_k, U_{l(k)}, U_{q(k)}, X_{q(k)}, Q) - \delta(\epsilon)$, $R_k^c + R_k^p < I(X_k; Y_{q(k)} | U_{l(k)}, U_{q(k)}, X_{q(k)}, Q) - \delta(\epsilon)$, $R_k^p + R_{l(k)}^c < I(X_k, U_{l(k)}; Y_{q(k)} | U_k, U_{q(k)}, X_{q(k)}, Q) - \delta(\epsilon)$, and $R_k^c + R_k^p + R_{l(k)}^c < I(U_k, X_k, U_{l(k)}; Y_{q(k)} | U_{q(k)}, X_{q(k)}, Q) - \delta(\epsilon)$ are satisfied, respectively, where $\delta(\epsilon)$ is a functions that tends to zero as $\epsilon \to 0$.

Finally, by substituting $R_k^p = R_k - R_k^c$ and using Fourier-Motzkin we obtain Lemma 1. ∎

## A. Proof of Theorem 1

We show that the rate regions in Theorem 1 and Lemma 1 are equal. For that, we use the same approach as in [10].

Let $\mathcal{P}$ be the set of pmfs on $\mathcal{Q} \times \mathcal{U}_1 \times \mathcal{X}_1 \times \mathcal{U}_2 \times \mathcal{X}_2 \times \mathcal{U}_3 \times \mathcal{X}_3$ that factor as $p(q)p(u_1, x_1|q)p(u_2, x_2|q)p(u_3, x_3|q)$. Fix an input pmf $P \in \mathcal{P}$ and let $\mathcal{R}^o(P)$ be the achievable rate region from Lemma 1 evaluated for $P$. For a fixed $\kappa \in \mathcal{K}$, this rate region is given below in explicit form:

$$R_\kappa < B_\kappa \tag{7}$$
$$R_{q(\kappa)} < B_{q(\kappa)} \tag{8}$$
$$R_{l(\kappa)} < B_{l(\kappa)} \tag{9}$$
$$R_\kappa < A_\kappa + C_{q(\kappa)} \tag{10}$$
$$R_{q(\kappa)} < A_{q(\kappa)} + C_{l(\kappa)} \tag{11}$$
$$R_{l(\kappa)} < A_{l(\kappa)} + C_\kappa \tag{12}$$
$$R_\kappa + R_{q(\kappa)} < A_\kappa + D_{q(\kappa)} \tag{13}$$
$$R_{q(\kappa)} + R_{l(\kappa)} < A_{q(\kappa)} + D_{l(\kappa)} \tag{14}$$
$$R_{l(\kappa)} + R_\kappa < A_{l(\kappa)} + D_\kappa \tag{15}$$
$$R_\kappa + R_{q(\kappa)} < A_\kappa + C_{q(\kappa)} + C_{l(\kappa)} \tag{16}$$
$$R_{q(\kappa)} + R_{l(\kappa)} < A_{q(\kappa)} + C_{l(\kappa)} + C_\kappa \tag{17}$$
$$R_{l(\kappa)} + R_\kappa < A_{l(\kappa)} + C_\kappa + C_{q(\kappa)} \tag{18}$$
$$R_\kappa + R_{q(\kappa)} + R_{l(\kappa)} < A_\kappa + C_{q(\kappa)} + D_{l(\kappa)} \tag{19}$$
$$R_{q(\kappa)} + R_{l(\kappa)} + R_\kappa < A_{q(\kappa)} + C_{l(\kappa)} + D_\kappa \tag{20}$$
$$R_{l(\kappa)} + R_\kappa + R_{q(\kappa)} < A_{l(\kappa)} + C_\kappa + D_{q(\kappa)} \tag{21}$$
$$R_\kappa + R_{q(\kappa)} + R_{l(\kappa)} < C_\kappa + C_{q(\kappa)} + C_{l(\kappa)} \tag{22}$$
$$2R_\kappa + R_{q(\kappa)} + R_{l(\kappa)} < A_\kappa + C_{q(\kappa)} + C_{l(\kappa)} + D_\kappa \tag{23}$$
$$2R_{q(\kappa)} + R_{l(\kappa)} + R_\kappa < A_{q(\kappa)} + C_{l(\kappa)} + C_\kappa + D_{q(\kappa)} \tag{24}$$
$$2R_{l(\kappa)} + R_\kappa + R_{q(\kappa)} < A_{l(\kappa)} + C_\kappa + C_{q(\kappa)} + D_{l(\kappa)}. \tag{25}$$

Then, $\mathcal{R}^o = \bigcup_{P \in \mathcal{P}} R^o(P)$ is the rate region in Lemma 1 for a fixed $f$. Further, define $\tilde{\mathcal{R}}^c(P)$ as the set of all rate tuples $(R_1, R_2, R_3)$ satisfying (7) to (9) and (13) to (25), and $\mathcal{R}^c(P)$ as the set of all rate tuples $(R_1, R_2, R_3)$ satisfying (7) to (9), (13) to (15) and (19) to (25). Let $\tilde{\mathcal{R}}^c = \bigcup_{P \in \mathcal{P}} \tilde{\mathcal{R}}^c(P)$ and $\mathcal{R}^c = \bigcup_{P \in \mathcal{P}} \mathcal{R}^c(P)$. Thus, $\mathcal{R}^c$ is the rate region in Theorem 1 for a fixed $f$. Our goal is to show that $\mathcal{R}^c = \mathcal{R}^o$.

Obviously, $\mathcal{R}^o \subseteq \tilde{\mathcal{R}}^c \subseteq \mathcal{R}^c$ since, for every $P \in \mathcal{P}$, $\mathcal{R}^o(P) \subseteq \tilde{\mathcal{R}}^c(P) \subseteq \mathcal{R}^c(P)$. Thus, for equality to hold, we have to show that $\mathcal{R}^c \subseteq \tilde{\mathcal{R}}^c \subseteq \mathcal{R}^o$. We first show that the second inclusion holds and proceed with a distinction of cases to show that, for every $P \in \mathcal{P}$, $\tilde{\mathcal{R}}^c(P) \subseteq \mathcal{R}^o$. First, if $R_k < A_k + C_{q(k)}$ for all $k \in \mathcal{K}$, then, obviously, $\tilde{\mathcal{R}}^c(P) \subseteq \mathcal{R}^o(P)$. Next, consider the case

$$R_\kappa \geq A_\kappa + C_{q(\kappa)}. \tag{26}$$

We construct a new input pmf $P_\kappa \in \mathcal{P}$ from $P$ that results in a $\mathcal{R}^o(P_\kappa)$ such that $\tilde{\mathcal{R}}^c(P) \subseteq \mathcal{R}^o(P_\kappa)$. For this inclusion to hold, we have to show that every inequality in $\mathcal{R}^o(P_\kappa)$ is implied by $\tilde{\mathcal{R}}^c(P)$ and (26).

Let $P_\kappa = \sum_{u_\kappa \in \mathcal{U}_\kappa} P$ be the marginal pmf of $P$ where $U_\kappa$ has been marginalized out and set $U_\kappa = \emptyset$. Then, $\mathcal{R}^o(P_\kappa)$ consists of all nonnegative rate tuples $(R_1, R_2, R_3)$ such that

$$R_\kappa < B_\kappa \tag{27}$$
$$R_{q(\kappa)} < B^*_{q(\kappa)} \tag{28}$$
$$R_{q(\kappa)} < A^*_{q(\kappa)} + C_{l(\kappa)} \tag{29}$$
$$R_{l(\kappa)} < B_{l(\kappa)} \tag{30}$$
$$R_\kappa + R_{l(\kappa)} < A_{l(\kappa)} + D_\kappa \tag{31}$$
$$R_{q(\kappa)} + R_{l(\kappa)} < A^*_{q(\kappa)} + D_{l(\kappa)} \tag{32}$$
$$R_\kappa + R_{q(\kappa)} + R_{l(\kappa)} < D_\kappa + A^*_{q(\kappa)} + C_{l(\kappa)} \tag{33}$$

where $A^*_{q(\kappa)} = I(X_{q(\kappa)}; Y_{l(\kappa)} | U_{q(\kappa)}, X_{l(\kappa)}, Q)$,
$B^*_{q(\kappa)} = I(X_{q(\kappa)}; Y_{l(\kappa)} | X_{l(\kappa)}, Q)$.

Equations (27), (30), and (31) are the same as (7), (9), and (15), respectively. From (6), (13) and (26), we obtain $R_{q(\kappa)} < I(U_{q(\kappa)}; Y_{l(\kappa)} | X_{l(\kappa)}, Q) \leq B^*_{q(\kappa)}$; from (16) and (26), we obtain $R_{q(\kappa)} < C_{l(\kappa)} \leq A^*_{q(\kappa)} + C_{l(\kappa)}$; from (19) and (26), we obtain $R_{q(\kappa)} + R_{l(\kappa)} < D_{l(\kappa)} \leq D_{l(\kappa)} + A^*_{q(\kappa)}$; and, finally, from (23) and (26), we obtain $R_\kappa + R_{q(\kappa)} + R_{l(\kappa)} < C_{l(\kappa)} + D_\kappa \leq C_{l(\kappa)} + D_\kappa + A^*_{q(\kappa)}$.

The proof for the cases $R_{q(\kappa)} < A_{q(\kappa)} + C_{l(\kappa)}$ and $R_{l(\kappa)} < A_{l(\kappa)} + C_\kappa$ follow exactly along the same lines due to symmetry. Hence, for every $P \in \mathcal{P}$, $\tilde{\mathcal{R}}^c(P) \subseteq \mathcal{R}^o(P) \cup \left(\bigcup_{k \in \mathcal{K}} \mathcal{R}^o(P_k)\right)$ with $P_k = \sum_{u_k \in \mathcal{U}_k} P$. Thus, $\tilde{\mathcal{R}}^c = \mathcal{R}^o$.

Next, we show that $\mathcal{R}^c \subseteq \tilde{\mathcal{R}}^c$ using similar steps as before. First, if $R_k + R_{q(k)} < A_k + C_{q(k)} + C_{l(k)}$ for all $k \in \mathcal{K}$, then $\mathcal{R}^c(P) \subseteq \tilde{\mathcal{R}}^c(P)$. Further, consider the case

$$R_\kappa + R_{l(\kappa)} \geq A_{l(\kappa)} + C_\kappa + C_{q(\kappa)}. \tag{34}$$

Let $P_\kappa$ be as before. Then, $\tilde{\mathcal{R}}^c(P_\kappa)$ consists of all nonnegative rate tuples $(R_1, R_2, R_3)$ such that

$$R_\kappa < B_\kappa \tag{35}$$
$$R_{q(\kappa)} < B^*_{q(\kappa)} \tag{36}$$
$$R_{l(\kappa)} < B_{l(\kappa)} \tag{37}$$
$$R_\kappa + R_{l(\kappa)} < A_{l(\kappa)} + D_\kappa \tag{38}$$
$$R_{q(\kappa)} + R_{l(\kappa)} < A^*_{q(\kappa)} + D_{l(\kappa)} \tag{39}$$
$$R_\kappa + R_{q(\kappa)} < B_\kappa + A^*_{q(\kappa)} + C_{l(\kappa)} \tag{40}$$
$$R_\kappa + R_{q(\kappa)} + R_{l(\kappa)} < D_\kappa + A^*_{q(\kappa)} + C_{l(\kappa)} \tag{41}$$

Equations (35), (37), and (38) are the same as (7), (9), and (15), respectively. From (6), (21) and (34), we obtain $R_{q(\kappa)} < I(U_{q(\kappa)}; Y_{l(\kappa)} | X_{l(\kappa)}, Q) \leq B^*_{q(\kappa)}$; from (25) and (34), we obtain $R_{q(\kappa)} + R_{l(\kappa)} < D_{l(\kappa)} \leq A^*_{q(\kappa)} + D_{l(\kappa)}$; from (6), (23) and (34), we obtain $R_\kappa + R_{q(\kappa)} < A_\kappa + D_\kappa - C_\kappa + C_{l(\kappa)} - A_{l(\kappa)} \leq B_\kappa + A^*_{q(\kappa)} + C_{l(\kappa)}$; and, finally, from (15), (22) and (34), we obtain $R_\kappa + R_{q(\kappa)} + R_{l(\kappa)} < C_{l(\kappa)} + D_\kappa \leq D_\kappa + A^*_{q(\kappa)} + C_{l(\kappa)}$.

The proof for the remaining cases $R_{q(\kappa)} + R_\kappa < A_\kappa + C_{q(\kappa)} + C_{l(\kappa)}$ and $R_{l(\kappa)} + R_{q(\kappa)} < A_{q(\kappa)} + C_{l(\kappa)} + C_\kappa$ is similar. Hence, $\mathcal{R}^c(P) \subseteq \tilde{\mathcal{R}}^c(P) \cup \left(\bigcup_{k \in \mathcal{K}} \tilde{\mathcal{R}}^c(P_k)\right)$. Thus, $\mathcal{R}^c = \tilde{\mathcal{R}}^c = \mathcal{R}^o$.

Finally, the cardinality bounds follow from a standard Carathéodory type argument [9, Appendix C]. ∎

*B. Proof of Corollary 2*

Set $U_k = X_k$, $k \in \mathcal{K}$, in Lemma 1. This implies $R_k^p = 0$. Observe that $\mathcal{E}_{q(k),1}$ and $\mathcal{E}_{q(k),3}$ do not contribute to the overall error probability of user $q(k)$. This is because they are associated with the decoding error for the "private" part of user $k$'s message. Since, here, all information is carried by $U_k$, the rate constraints due to $\mathcal{E}_{q(k),1}$ and $\mathcal{E}_{q(k),3}$ are no longer applicable and can be omitted. Finally, the cardinality bound on $\mathcal{Q}$ follows from a standard Carathéodory type argument [9, Appendix C]. ∎

## IV. GAUSSIAN MWRC WITH AF RELAYING

Consider the Gaussian MWRC with AF relaying. It is defined by $Y_0 = \sum_{k \in \mathcal{K}} X_k + Z_0$, $Z_0 \sim \mathcal{N}(0, N_0)$, and $Y_k = X_0 + Z_k$, $Z_k \sim \mathcal{N}(0, N_k)$, for $k \in \mathcal{K}$, with average power constraint $P_k$ on $X_k$, $k \in \mathcal{K} \cup \{0\}$, and a linear relaying function $X_0 = \alpha Y_0$, where $\alpha$ is a normalization factor chosen such that the transmit power constraint at the relay is met. Theorem 1 can be adapted to Gaussian channels using the standard procedure [9, Chapter 3]. We evaluate it for Gaussian inputs $U_k \sim \mathcal{N}(0, \bar{\lambda}_k)$ and $X_k = U_k + V_k$, with $V_k \sim \mathcal{N}(0, \lambda_k)$, where $\lambda_k$ is chosen such that the interference to noise ratio is one (or as close to one as possible), i.e., $\lambda_k = \min\left\{1, \frac{N_0 + \alpha^{-2} N_{l(k)}}{P_k}\right\}$. This power allocation was proposed in [11].

Figure 2 shows the achievable sum rate of HK coding compared to SND and IAN for $P_0 = P$, $N_0 = N$, and $P_k = |h_k|^2 P$ and $N_k = \frac{N}{|h_k|^2}$, $k \in \mathcal{K}$, with $h = [0.1\ 0.5\ 0.9]$. SND and IAN results were obtained from Corollaries 1 and 2, respectively.[1] First of all, observe that HK and IAN perform virtually the same up to approximately $15\,\text{dB}$. This can be explained by the small $P$ to $N$ ratio which results in a power allocation $\lambda_k \approx 1$, $k \in \mathcal{K}$. Furthermore, HK coding achieves higher sum rates than SND for all $\frac{P}{N}$ up to approximately $40\,\text{dB}$. However, for some signal-to-noise ratios (SNRs), HK coding is outperformed by SND, but it still dominates in the high SNR regime. Also, by choosing $\lambda_k = 0$ for all $k \in \mathcal{K}$ for those SNRs where SND dominates, the sum rate achieved by HK coding is always greater or equal than the rate achieved by SND.

## V. DISCUSSION

In [12], the HK region for the $K$-user cyclic IC is computed. In some cases, such as Gaussian channels with linear relaying functions, each terminal can subtract its self-interference easily from its received signal. Then, the resulting channel is equivalent to a cyclic IC and the results from [12] are applicable. However, the results presented here cover the much more general case of arbitrary instantaneous relaying functions and DM channels. For this class of channels, we derived, to the best of our knowledge, the largest achievable rate region known to date for restricted encoders. Now, a natural question to ask would be how much could be gained by using the feedback information present at the encoders. This topic will be dealt with in future research.

---

[1]Alternatively, SND and IAN results could be obtained by setting $\lambda_k = 0$ and $\lambda_k = 1$, respectively, for all $k \in \mathcal{K}$.

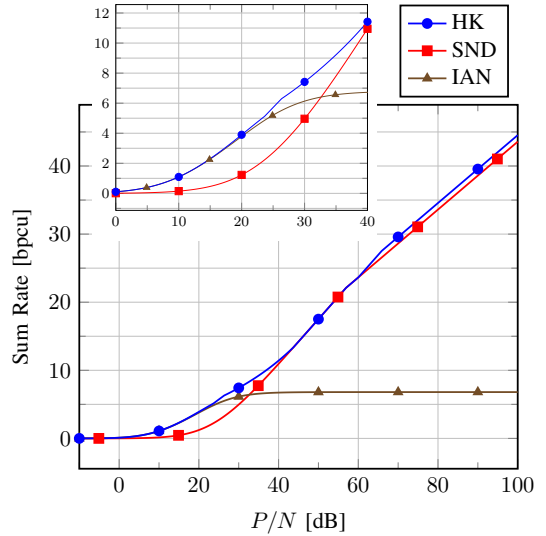

Fig. 2. Achievable sum rates in the Gaussian MWRC with AF relaying and channels $h = [0.1\ 0.5\ 0.9]$; 1) Han-Kobayashi (HK) coding, 2) simultaneous non-unique decoding (SND), and 3) treating interference as noise (IAN) plotted as a function of $\frac{P}{N}$.